\documentclass[prl,reprint,superscriptaddress,nofootinbib,twocolumn,nolongbibliography]{revtex4-2}

\usepackage[linktoc=page,colorlinks,urlcolor=blue,citecolor=blue,linkcolor=blue]{hyperref}
\usepackage{amssymb,amsmath}
\usepackage{graphicx}
\usepackage{natbib}
\usepackage{mathrsfs}
\usepackage{overpic}
\usepackage{gensymb}
\usepackage{lineno}
\usepackage{color}
\usepackage{upgreek}
\usepackage[dvipsnames]{xcolor}
\usepackage[letterpaper,textwidth=7in,top=.75in,bottom=.75in]{geometry}
\usepackage{textcomp}

\usepackage{pdfpages}
\makeatletter
\AtBeginDocument{\let\LS@rot\@undefined}
\makeatother

\newcommand{\snl}{Sandia National Laboratories, Albuquerque, New Mexico 87185, USA}
\newcommand{\cint}{Center for Integrated Nanotechnologies, Sandia National Laboratories, Albuquerque, New Mexico 87123, USA}

\begin{document}

\title{Fault Localization in a Microfabricated Surface Ion Trap using Diamond Nitrogen-Vacancy Center Magnetometry}

\date{\today}

\author{Pauli Kehayias}
\altaffiliation{Present address: MIT Lincoln Laboratory, Lexington, Massachusetts, 02421, USA}
\affiliation{\snl}

\author{Matthew A. Delaney}
\affiliation{\snl}

\author{Raymond A. Haltli}
\affiliation{\snl}

\author{Susan M. Clark}
\affiliation{\snl}

\author{Melissa C. Revelle}
\affiliation{\snl}

\author{Andrew M. Mounce}
\email{amounce@sandia.gov}
\affiliation{\cint}

\begin{abstract}
As quantum computing hardware becomes more complex with ongoing design innovations and growing capabilities, the quantum computing community needs increasingly powerful techniques for fabrication failure root-cause analysis.  This is especially true for trapped-ion quantum computing.
As trapped-ion quantum computing aims to scale to thousands of ions, the electrode numbers are growing to several hundred with likely integrated-photonic components also adding to the electrical and fabrication complexity, making faults even harder to locate.
In this work, we used a high-resolution quantum magnetic imaging technique, based on nitrogen-vacancy (NV) centers in diamond, to investigate short-circuit faults in an ion trap chip. We imaged currents from these short-circuit faults to ground and compared to intentionally-created faults, finding that the root-cause of the faults was failures in the on-chip trench capacitors.  This work, where we exploited the performance advantages of a quantum magnetic sensing technique to troubleshoot a piece of quantum computing hardware, is a unique example of the evolving synergy between emerging quantum technologies to achieve capabilities that were previously inaccessible. 
\end{abstract}
\maketitle

\section{Introduction}
Quantum computing is a promising field investigating new computing capabilities using qubits and quantum algorithms.  Superconducting qubits, Rydberg atoms, quantum dots, solid-state defect centers, photons, and trapped ions are among the candidate technologies being investigated for quantum computing.
Trapped-ion qubits are a leading technology for achieving a scalable quantum computer, due to ion qubits having long coherence times, optical and microwave state manipulation, controllable and reconfigurable interactions between ions, and the possibility to use a room-temperature apparatus. Microfabricated ion trap chips are an appealing approach for creating the electrostatic and radiofrequency (RF) potentials needed for scalable trapped-ion quantum computing, as they offer the possibility of repeatable large-scale fabrication and the ability to route electronic and photonic paths under the trap surface \cite{winelandSurfaceTrap2006, stickIonChipTrap2006}. As ion trap chips become more sophisticated (with integrated photonics, photodiodes, electrodes, and electrical routing), traditional electronics failure analysis (FA) techniques have a growing challenge to keep up with their increasing complexity \cite{HIFAreport, asmFAdeskRef, wagnerFAtechniques, ionTrapEDFAS, edfasRoadmap2023}. In particular, many FA techniques that rely on optical access  may struggle to detect faults deep within ion trap chips with high-density layouts and a high metal density.

\begin{figure*}[ht]
\begin{center}
\begin{overpic}[width=0.95\textwidth]{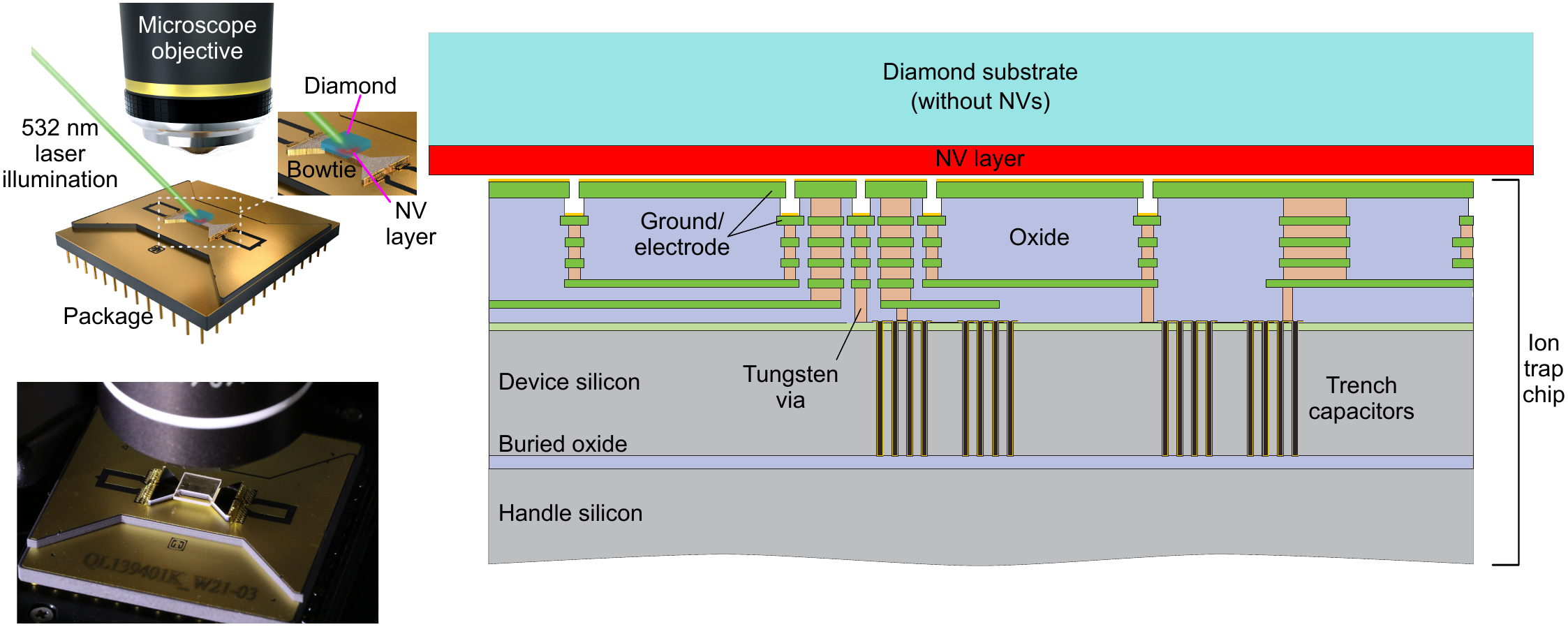}
\put(1,39){\textsf{\Large a}}
\put(1,17){\textsf{\Large b}}
\put(27,39){\textsf{\Large c}}
\end{overpic}
\end{center}
\caption{\label{fig1}
(a) Schematic drawing of the diamond on top of an ion trap chip in a QDM setup. A 532 nm laser beam illuminates the NV layer, and the red NV fluorescence is collected by a microscope objective and imaged onto a camera with an optical microscope (not shown).
(b) A photo of an example ion trap chip in a QDM setup. 
(c) Cross-section drawing of the Roadrunner ion trap chip, showing the metal and insulating layers, tungsten vias, trench capacitors, and the device and handle silicon. For the MFI experiments, the diamond sample (containing an NV layer) is mounted on top. 
}
\end{figure*}

Magnetic field imaging (MFI) is an attractive nondestructive fault-localization technique for locating and diagnosing short-circuit faults \cite{orozcoBookCh}. Using an MFI instrument, one maps the magnetic field emissions from currents within a device, then infers the fault location by interpreting the current paths from the resulting magnetic images. MFI is appealing because most microfabrication materials (metals, semiconductors, and insulators) are transparent to magnetic fields, making it possible to look deep into an otherwise-opaque device independent of what materials are nearby. In addition, MFI can yield information about internal current paths, which is not possible with many other FA techniques. 
One instrument in particular, the quantum diamond microscope (QDM) \cite{QDM1ggg, edlynQDMreview, tetienneQDMreview}, has shown recent promise for high-resolution MFI for electronics failure analysis and interrogation \cite{qdmFPGA, NV555, NVAPAM, NVphotovoltaics, nvTIVA, hollenbergCurrentImg, fpgaBackThin, edlynIstfa2021}. By putting a diamond sample with a thin surface layer of magnetically-sensitive nitrogen-vacancy (NV) color centers on top of a device (in our case, an ion trap chip), illuminating the NVs with green laser light, and imaging the NV fluorescence with an optical microscope, we can extract the magnetic field in every pixel (Fig.~\ref{fig1}a-b). Because QDM setups yield micron-scale spatial resolution over a few-millimeter field of view, measure all pixels simultaneously without scanning, and excel at detecting shallow magnetic sources within a few microns of the diamond surface, they are an ideal choice for electronics interrogation.

In this paper, we describe our work using a QDM to locate short-circuit faults in an ion trap chip, leading to the discovery of the root-cause explanation for these faults.  We studied an ion trap chip of the Sandia Roadrunner design \cite{Revelle2024}  with several short-circuit faults (electrode shorts to ground).  Initial use of traditional test methods was unable to  narrow down the root cause from candidates such as  electrostatic discharge (ESD), scratches at the bond pads, scratches at the electrodes, conducting particles shorting the metal layers, and others. 
After imaging the magnetic fields from probe currents applied to the device, we found that every current path dissipated at a trench capacitor, indicating that the integrated trench capacitors were most likely the cause.
To confirm this, we added intentional faults (overloaded capacitors, scratches on the bond pads, and scratches on the electrodes) and compared the magnetic images of the ``natural" and ``intentional" faults to validate this finding.

This work informs future ion trap chip designs and fabrication recipes, as we now know to give extra scrutiny to the trench capacitor fabrication steps.  Additionally, we can avoid the potential problems introduced by trench capacitors by replacing them with external chiplet trench capacitors that can be tested separately before assembly.  This result also confirms that the other possible failure mechanisms (e.g.~assembly) don't need further scrutiny at this time. Furthermore, as the QDM gains traction as a novel tool for electronics interrogation, this work represents a unique application direction. In addition to troubleshooting sophisticated next-generation integrated circuits (ICs), this quantum sensing approach can also solve problems for quantum computing hardware, supporting the development of computing technology that can surpass the current classical-computing limits.

\begin{figure*}[ht]
\begin{center}
\begin{overpic}[width=0.95\textwidth]{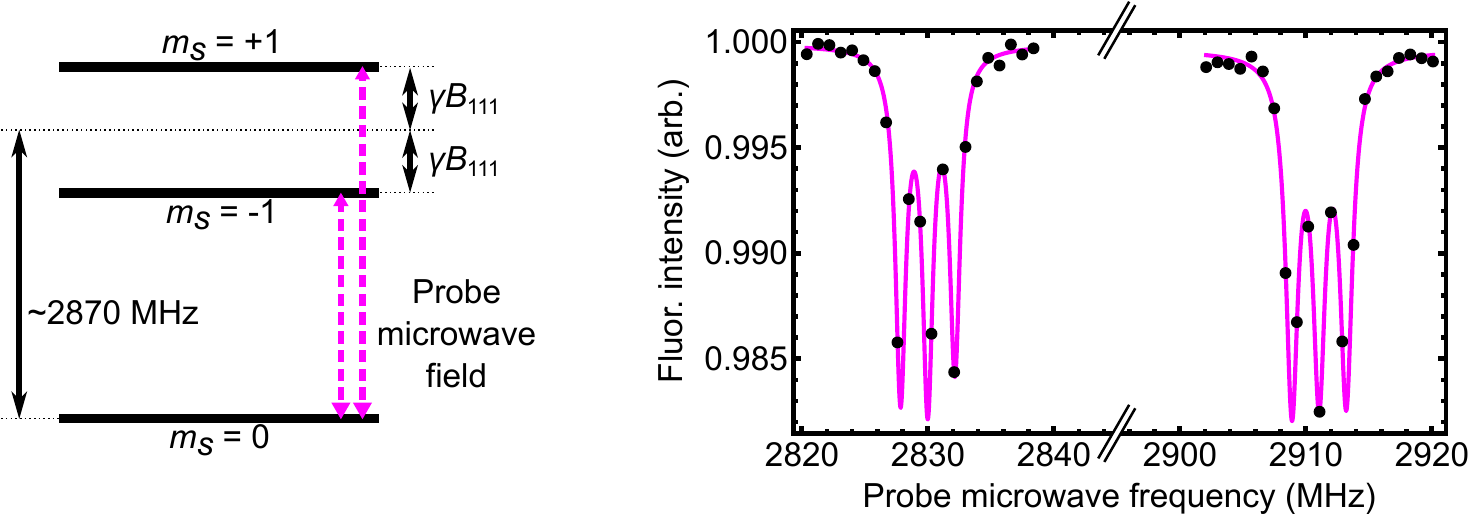}
\put(0,31){\textsf{\Large a}}
\put(43,31){\textsf{\Large b}}
\end{overpic}
\end{center}
\caption{\label{nvBgFig}
(a) An energy-level drawing of the NV ground-state magnetic sublevels. Given a magnetic field $B_{111}$ along the NV axis, the $m_s = \pm 1$ sublevels shift by $\pm \gamma B_{111}$ in frequency.
(b) An example single-pixel ODMR spectrum. We fit each lineshape to extract its center frequency, which yields the measured $B_{111}$ for that pixel. The lineshapes are split into three peaks, due to the NV $^{14}$N hyperfine interaction.
}
\end{figure*}

\section{Materials and methods}

\subsection{Ion trap chip specifications}

The ion trap chip tested in this work is the latest device from the Quantum Scientific Computing Open User Testbed (QSCOUT) program at Sandia National Laboratories, the Roadrunner trap \cite{Revelle2024}. 
Similar to other recently microfabricated traps from Sandia (such as the Peregrine and Phoenix traps \cite{phoenixPeregrine}), the Roadrunner ion trap chip has six metal layers for routing control electrode signals. It also has integrated trench capacitors to shield the applied DC voltages from the RF trapping potential, gold coating for improved ion qubit performance, and a junction to allow for ion re-ordering  (Fig.~\ref{fig1}c). 
The trap has a bowtie shape to improve the optical access for lasers addressing the ions, which constrains all of the bond pads to the edge of the trap~\cite{HOAManual}. 
On each edge of the trap there are 50 control electrode wire bonds, along with connections for the RF trapping voltage and ground. 
The control electrode routing to the edge of the trap (for wire bonding) occurs through the lower metal layers. Additionally, this routing connects to trench capacitors located in the device silicon. 
The device chip is attached to a custom 30~mm  102-connection bowtie package, where each electrode is linked to a pin in a 13$\times$13 array.

We detected nine electrode shorts to ground in the device, with resistances ranging from 60-160 $\ohm$. These shorts were discovered electrically, but the root-cause explanation remained unknown. Furthermore, the faults could be located anywhere along the paths between the bond pads and the electrodes, and these paths span millimeters in length and several conducting layers in the device. However, using a QDM to map the internal current paths with high spatial resolution and comparing to the device layout geometry, we can locate the faults, collect statistics on what components are failing, identify which elements in the fabrication process and/or design are responsible, and implement  future design changes to mitigate these failure mechanisms.

\subsection{NV magnetic microscopy apparatus}

Our QDM setup is a home-built optical fluorescence microscope, with a kinematic sample stage to mount devices for MFI measurements. The ion trap chip was mounted in a zero insertion force (ZIF) electrical socket on the sample stage, through which a probe current (typically $\pm$20 mA) was applied to the ion trap chip.  We placed a $4\times4\times0.5$ mm$^3$ diamond sample  with a 4 $\upmu$m sheet of magnetically-sensitive NV centers on top of the device (Fig.~\ref{fig1}). The NV electronic ground-state magnetic sublevels are sensitive to the magnetic field $B_{111}$ along the N-V axis (the [111] crystallographic direction) through the Zeeman effect. This leads to the $m_s = 0 $ to $m_s = \pm1$ transition frequencies being shifted by $\pm \gamma B_{111}$, where $\gamma \approx 28$ GHz/T is the NV gyromagnetic ratio (Fig.~\ref{nvBgFig}a). We used 200-400 mW of 532 nm laser light to illuminate the NV layer from the side of the microscope objective, which pumps the NVs into the $m_s =0$ sublevel. A resonant microwave field from a nearby wire loop (not shown in Fig.~\ref{fig1}) transfers NV population from the $m_s =0$ ``bright state" to the $m_s =\pm1$ ``dark state" sublevels, reducing the NV fluorescence intensity. Scanning the microwave frequency across the NV resonance frequencies and imaging the NV fluorescence with the microscope, we collect an optically-detected magnetic resonance (ODMR) spectrum for each camera pixel (Fig.~\ref{nvBgFig}b). We fit the ODMR spectra for each pixel, extract the lineshape center frequencies, and obtain a resulting $B_{111}$ map detected by the NV layer \cite{QDM1ggg, edlynQDMreview}.  

We used a pair of permanent magnets (not shown in Fig.~\ref{fig1}) to apply a $1.4~\mathrm{mT}$ bias magnetic field along the N-V axis. For each field of view, we measured the magnetic field maps for equal and opposite probe currents ($\pm$20 mA), then subtracted the results. This yields the magnetic field map from the probe current only, excluding the bias magnetic field contribution.
Multiple magnetic images were stitched together after measuring the device in several locations. Our magnetic images are 2.03 $\times$ 1.27 mm$^2$ in size, and typically took 20 minutes to acquire.

\begin{figure*}[ht]
\begin{center}
\begin{overpic}[width=0.95\textwidth]{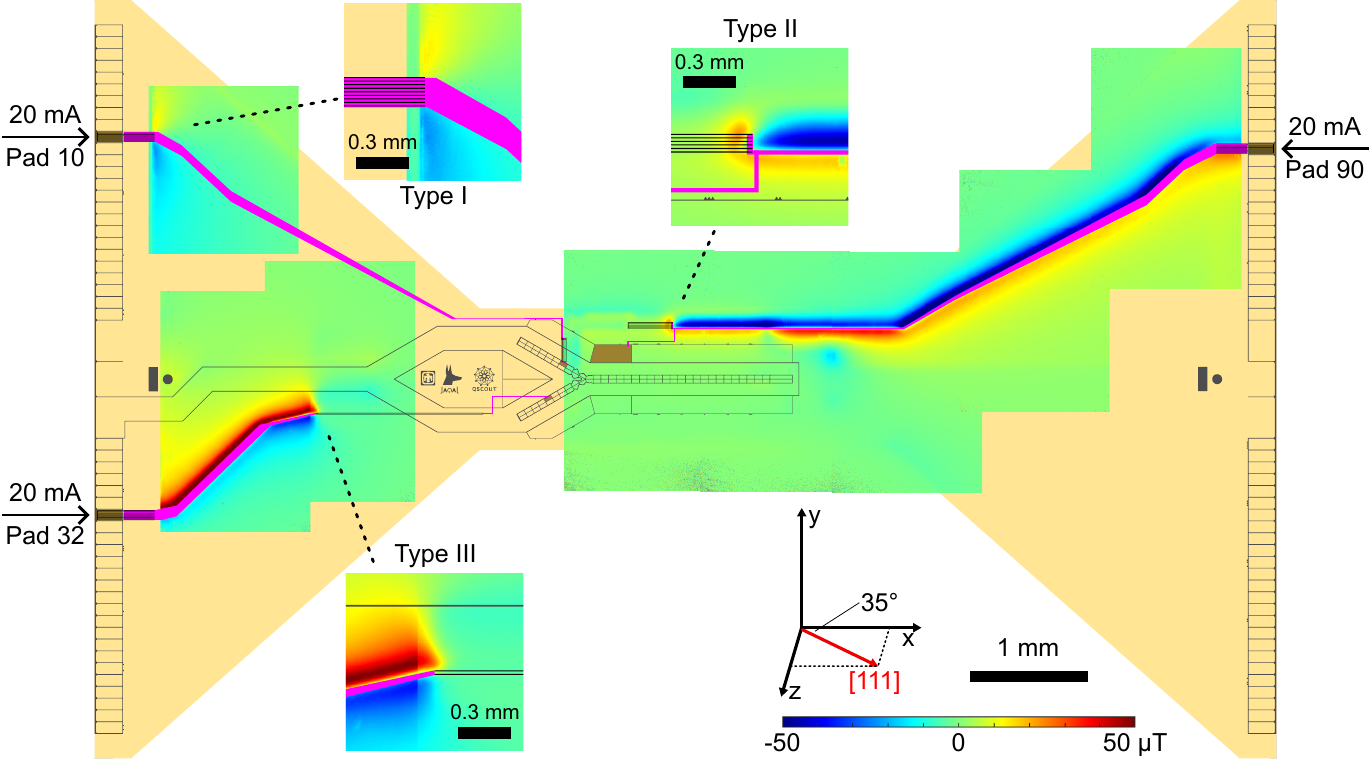}
\end{overpic}
\end{center}
\caption{\label{resultsFig1}
Magnetic field maps (measured along the NV [111] direction) for natural shorts in the ion trap chip, superimposed on the top metal layer. The magenta traces show the paths between the bond pads and the trapping-region electrodes, which the applied currents follow until they short to ground. We include three examples faults (one for each category): one near the bond pads (Pad 10), one in the center isthmus (Pad 90), and one at a capacitor bank (Pad 32).  The insets show the $1\times1~\mathrm{mm}^2$ fault locations enlarged, indicating that the faults occur at trench capacitor edges (black).
}
\end{figure*}

\begin{figure*}[ht]
\begin{center}
\begin{overpic}[width=0.95\textwidth]{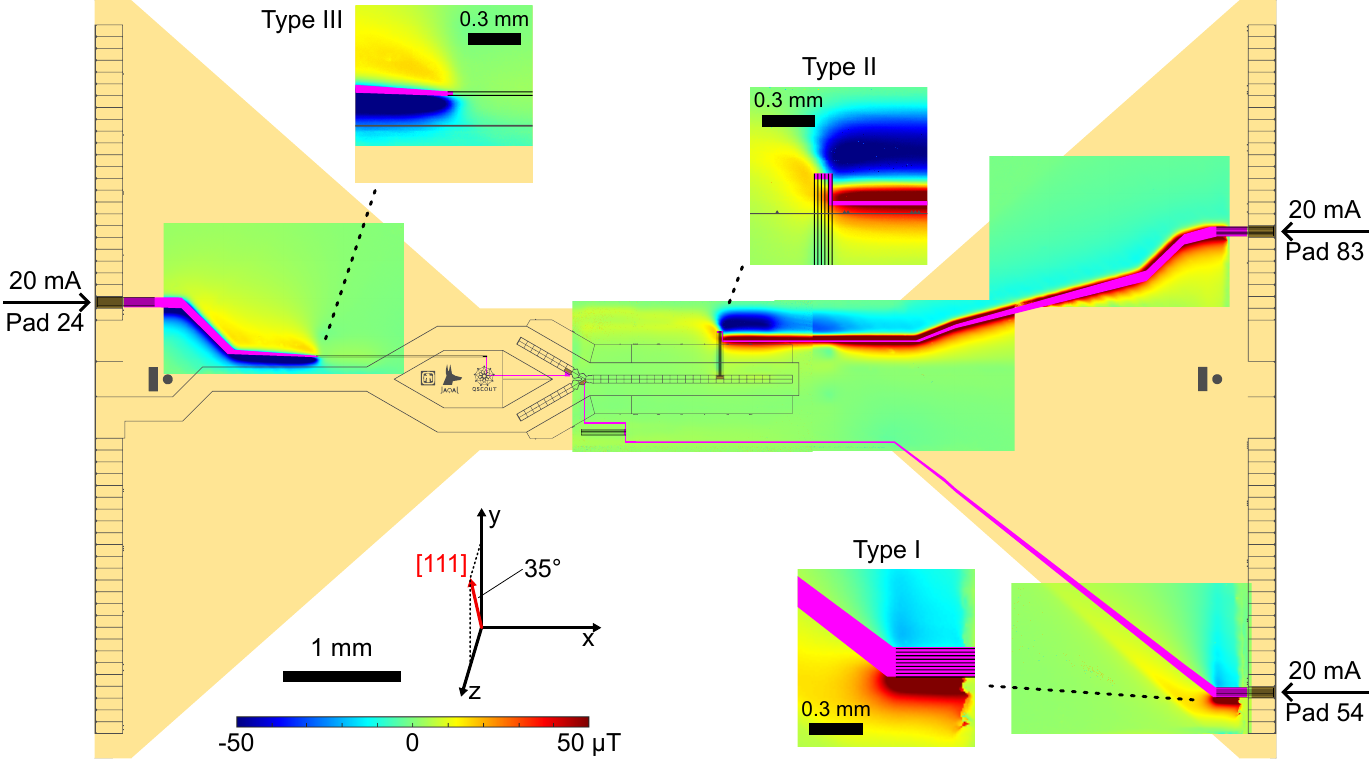}
\end{overpic}
\end{center}
\caption{\label{resultsFig2}
Magnetic field maps for intentionally overloaded capacitors in the ion trap chip, superimposed on the top metal layer. The magenta traces show the paths between the bond pads and the trapping-region electrodes, which the applied currents follow until they short to ground. We include three example faults (one for each category): one near the bond pads (Pad 54), one in center isthmus (Pad 83), and one at a capacitor bank (Pad 24). The similarities here to Fig.~\ref{resultsFig1} supports the hypothesis that trench capacitors (black) are the culprit.
}
\end{figure*}

\begin{figure*}[ht]
\begin{center}
\begin{overpic}[width=0.95\textwidth]{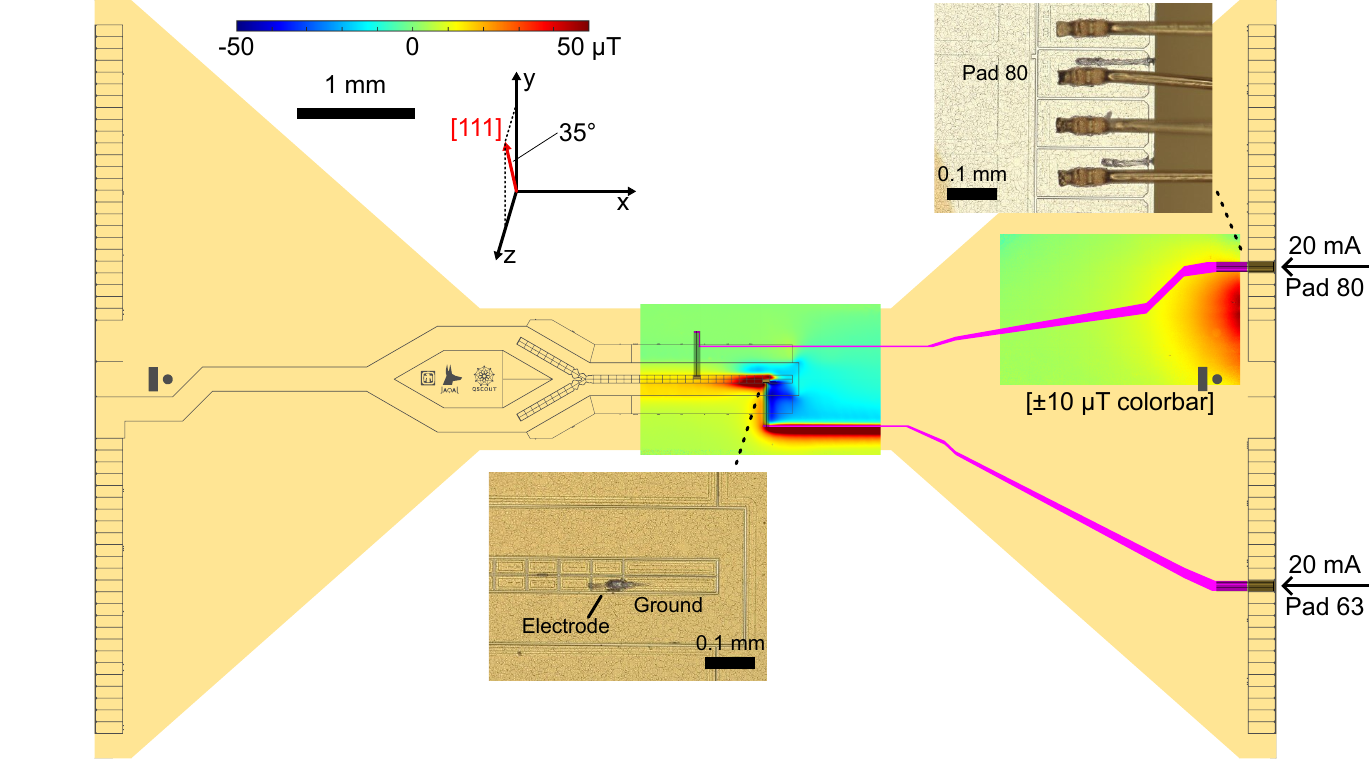}
\end{overpic}
\end{center}
\caption{\label{resultsFig3}
Magnetic field maps for intentional scratches on the top metal layer of the ion trap chip, superimposed on the top metal layer. The magenta traces show the paths between the bond pads and the trapping-region electrodes, which the applied currents follow until they short to ground. We include two examples: a bond-pad scratch (Pad 80, shown with a zoomed-in $\pm10~\upmu$T colorbar) and an electrode scratch (Pad 63). The inset photos show the scratches, including one on Pad 78 that was not measured. 
Comparing to the characteristics seen in Fig.~\ref{resultsFig1}, these magnetic images indicate that the natural faults were probably not caused by scratches on the top metal layer. 
}
\end{figure*}

\section{Results}
To diagnose the shorts in our ion trap devices, we used our QDM instrument to magnetically image the currents flowing through the device, mapping the current flow from the bond pads to the current termination points.
After magnetically imaging the ``natural" shorts (i.e~shorts caused during manufacturing), we found that all of them were located at trench capacitors, suggesting trench capacitors as the failure point. We confirmed this hypothesis by creating ``intentional" shorts, either by overloading previously non-shorted capacitors or by scratching bond pads and electrodes to cause shorts to ground. Comparing the magnetic images of natural and intentional shorts, we confirmed trench-capacitor failures as the root cause.

\subsection{Initial Testing: Natural Shorts}
Figure \ref{resultsFig1} shows magnetic images from three faults, overlaid on the ion trap chip. The magnetic images show the applied probe currents traveling along internal traces (magenta) towards their respective electrodes (brown) before shorting to ground and dissipating. Measuring magnetic images for the nine faults, we found these faults fell into three categories:
\begin{itemize}
    \item Type I: Five faults near the bond pads (e.g.~Pad 10). These occur at  trench capacitors (black) next to the electrodes, as illustrated by the inset.
    \item Type II: Three faults in the center isthmus (e.g.~Pad 90). These occur at  trench capacitors (black) in the center isthmus, as illustrated by the inset.
    \item Type III: One fault $\sim$2 mm from the bond pads (Pad 32). This occurs at a trench capacitor bank (black) along the path between the bond pad and the electrode.
\end{itemize}

As indicated by the insets in Fig.~\ref{resultsFig1}, the locations where the probe currents dissipate to ground suggest that trench capacitors are the culprit. To further investigate this finding, we introduced intentional short-circuit faults into the device, then measured their corresponding magnetic field images.

\subsection{Hypothesis Testing: Intentional Shorts (Overloaded Capacitors)}
To validate the hypothesis that trench capacitors are the root-cause explanation for the faults, we overloaded several other capacitors (connected to previously intact pins), causing them to short to ground, then measured their magnetic images for comparison. As seen in Fig.~\ref{resultsFig2}, the capacitor-overload intentional-short magnetic images resemble those of the natural shorts, with Pads 54, 83, and 24 resembling faults of Types I, II, and III, respectively. The similarities between Fig.~\ref{resultsFig1} and Fig.~\ref{resultsFig2} supports the trench capacitor  failure root-cause explanation.

\subsection{Hypothesis Testing: Intentional Shorts (Scratches)}
To rule out the other fault root-cause explanation possibilities, we also added additional faults by scratching the top metal layer with a probe tip (Fig.~\ref{resultsFig3}), again on previously intact pins. MFI above the bond pads was not attempted because of the bond wires in the way \cite{NV555}. However, after scratching Pad 80, the resulting magnetic field map shows the fringe of a magnetic feature to the right of the field of view. This differs from the natural Type I faults, where current dissipates more inland with stronger and more compact magnetic features.  Conversely, for the scratched electrode in the center isthmus (connected to Pad 63), the probe current traces the entire path before dissipating at the electrode itself. This differs from the natural Type II faults, where the current dissipates before reaching the electrode. Since there are no bond pads or electrodes $\sim$2 mm from the bond pads, Type III faults are unlikely to be reproduced by these intentional scratches.   Finally, we note that scratches on the top metal layer should be visible with an optical microscope, while shorts in a trench capacitor (in the deeper layers) should be invisible.  While it is difficult to create intentional shorts to ground due to conducting particles between the metal layers, these hypothetical shorts should occur in random locations, which is inconsistent with the fault/trench capacitor co-location. These arguments provide a convincing case that trench capacitor faults are responsible for the shorts in this ion trap chip.

\section{Discussion}
The above results lead us to conclude that the faults are due to the trench capacitors, ruling out the other possible explanations (e.g.~wire bond faults, packaging, and metal particle contamination during fabrication) and saving us from further analysis efforts pursuing solutions to those potential problems.  We identified several methods by which the trench capacitors could fail. One possibility is ESD in the packaging lab or silicon fabrication facility. Alternatively, the trench capacitor faults could be due to conducting particles landing on the device during the relevant trench capacitor fabrication steps. The investigation into distinguishing these possible explanations is ongoing.

Measuring a second ion trap chip of a different design (a Peregrine trap \cite{phoenixPeregrine}) with natural shorts spanning $\sim$100 $\ohm$ to $\sim$1 M$\ohm$, we also found that its faults were also located at its trench capacitors \cite{suppl}. Since two different ion trap chip designs made in different fabrication runs having different short resistances \textit{have the same fault root-cause explanations}, this suggests a possible common underlying mechanism. Furthermore, the combination of these results informs  future ion trap chip designs to either avoid trench capacitors (replacing them with external chiplet trench capacitors that can be tested separately before assembly) or to spend additional effort troubleshooting the  fabrication process. 
Overall, these results inform future designs and fabrication attempts, enabling a more robust fabrication process for better yield.

Other attempts to pinpoint the exact fault locations, such as cross-sectioning the device using a focused ion beam (FIB) and looking for anomalies with a scanning electron microscope (SEM), were unsuccessful. This can be interpreted as an advantage of our high-resolution QDM MFI approach, since MFI was able to locate the faults when cross-sectioning could not (even when the locations were already known).

\section{Conclusions}
In this work, we used a QDM apparatus for high-resolution MFI to locate short-circuit faults in an ion trap chip, leading to the conclusion that the faults were  caused by  trench capacitors.  By intentionally overloading other capacitors in the same device, and by scratching other bond pads and electrodes, we were able to confirm that trench capacitors were the root cause.  This  rules out other possible causes, such as  mechanical damage in the post-fabrication assembly processes.  
More generally, this work shows that, in addition to troubleshooting multilayer and multi-die ICs for sophisticated next-generation electronic components, a QDM can also apply its quantum sensing advantages to solve problems for quantum computing hardware as well \cite{ionTrapEDFAS, edfasRoadmap2023}.

Solving FA problems is especially important for next-generation ion trap chips with increasingly sophisticated designs, such as on-chip single-photon avalanche diodes (SPADs) and on-chip photonic waveguides for laser routing \cite{mitllSPADs, Niffenegger2020, sandiaSPADs, ivoryPRX2021, homeQCCD, nistSNSPDs, vermaSNSPDs2023}. 
We note that our ion trap chip fabrication is on a small production scale, where five good devices are sufficient for the intended experiments.  This means that, unless the failure rate is substantial (approaching $100\%$), we can usually tolerate some failed devices, and we can sometimes avoid having to solve a low-yield problem with a brute-force fabrication approach.  However, if this were a large-scale commercial fabrication effort, then our QDM failure analysis method is even more useful for locating where faults occur in the devices. In fact, commercial trapped-ion quantum computing entities are more motivated to use this approach to maximize their yield. Just as FA development is needed to match increasingly-complex electronics fabrication, the same is also true for quantum computing hardware if this technology is to be realistically scalable.

Quantum mechanics is transitioning from being an observed physics phenomenon (the ``first quantum revolution") to being technologies where we use quantum systems to achieve outcomes that are impossible classically (the ``second quantum revolution") \cite{secondQuantumRevolution}. Both the QDM and ion trap chip are examples of such quantum technologies. We note that this work, where we exploit the advantages of a quantum sensor to troubleshoot a piece of quantum computing hardware, is a unique example of one quantum technology supporting another.  Finally, in addition to being useful for FA, we anticipate that a QDM can provide additional 2D and 3D electromagnetic characterization information for ion trap chips that may be otherwise unavailable. These include mapping the RF trapping potential ($\sim$50 MHz) and/or microwave gates fields (GHz-frequency \cite{dianaCaMWgates, wineland2011MWgates}, to validate that these agree with what is expected from simulation, ensure that other nearby components in the vacuum apparatus (such as the ion trap chip socket) are not significantly altering the ion trapping environment, measure the DC magnetic field gradients caused by magnetic components in the ion trap chip (e.g.~Kovar pins), and study electrical noise due to time-varying patch potentials on the electrode surfaces \cite{blattIonTrapPatchNoiseRMP}.

\section{Acknowledgements}
We thank D.~Stick and C.~Nordquist for helpful discussions, R. Johnson for the drawing in Fig.~\ref{fig1}a, and L.~Basso for the photograph in Fig.~\ref{fig1}b.  Sandia National Laboratories is a multi-mission laboratory managed and operated by National Technology and Engineering Solutions of Sandia, LLC, a wholly owned subsidiary of Honeywell International, Inc., for the DOE's National Nuclear Security Administration under contract DE-NA0003525. This work was funded, in part, by the Laboratory Directed Research and Development Program and performed, in part, at the Center for Integrated Nanotechnologies, an Office of Science User Facility operated for the U.S.~Department of Energy (DOE) Office of Science.  This material is also funded, in part, by the U.S. Department of Energy, Office of Science, Office of Advanced Scientific Computing Research Quantum Testbed Program.  This paper describes objective technical results and analysis. Any subjective views or opinions that might be expressed in the paper do not necessarily represent the views of the U.S. Department of Energy or the United States Government.

%

%\bibliography{biblio}

\clearpage


\section*{Supplementary material}

# Supplemental Material for "Fault Localization in a Microfabricated Surface Ion Trap using Diamond Nitrogen-Vacancy Center Magnetometry"

Pauli Kehayias,[1, *] Matthew A. Delaney,[1] Raymond A. Haltli,[1]
Susan M. Clark,[1] Melissa C. Revelle,[1] and Andrew M. Mounce[2, †]

[1]*Sandia National Laboratories, Albuquerque, New Mexico 87185, USA*
[2]*Center for Integrated Nanotechnologies, Sandia National Laboratories, Albuquerque, New Mexico 87123, USA*

## MFI MEASUREMENTS OF A PEREGRINE TRAP

We measured a second ion trap chip of the Peregrine design [S1], which also had naturally-occurring short-circuit faults to ground. From initial testing, we found that this device had two classes of shorts: low-resistance faults ($\sim$100 $\Omega$, similar to the Roadrunner device measured in the main text) and high-resistance faults ($\sim$0.1-1 M$\Omega$). Since these two ion trap chips have different designs and were manufactured in different fabrication runs, here we test whether our finding that the shorts are always in the trench capacitors still holds.

Figure S1 shows magnetic images for three low-resistance shorts (Pads 18, 62, and 68) and a high-resistance short (Pad 80). We applied a $\pm$50 µA test current for the high-resistance short (instead of $\pm$20 mA) to avoid the possibility of arcing through air. As seen in the overlays, all of the shorts are located at trench capacitors, which appear as black squares. This suggests that the trench capacitor faults persist over multiple ion trap chip designs and fabrication runs (meaning there could be a possible common underlying root-cause explanation), and that trench capacitor faults can manifest as high-resistance shorts as well.

## PEREGRINE BOWTIE LAYOUT

Figure S2 shows the Roadrunner ion trap chip top metal layer, with a $4 \times 4$ mm$^2$ diamond sample for scale.

---

* Present address: MIT Lincoln Laboratory, Lexington, Massachusetts, 02421, USA
† amounce@sandia.gov

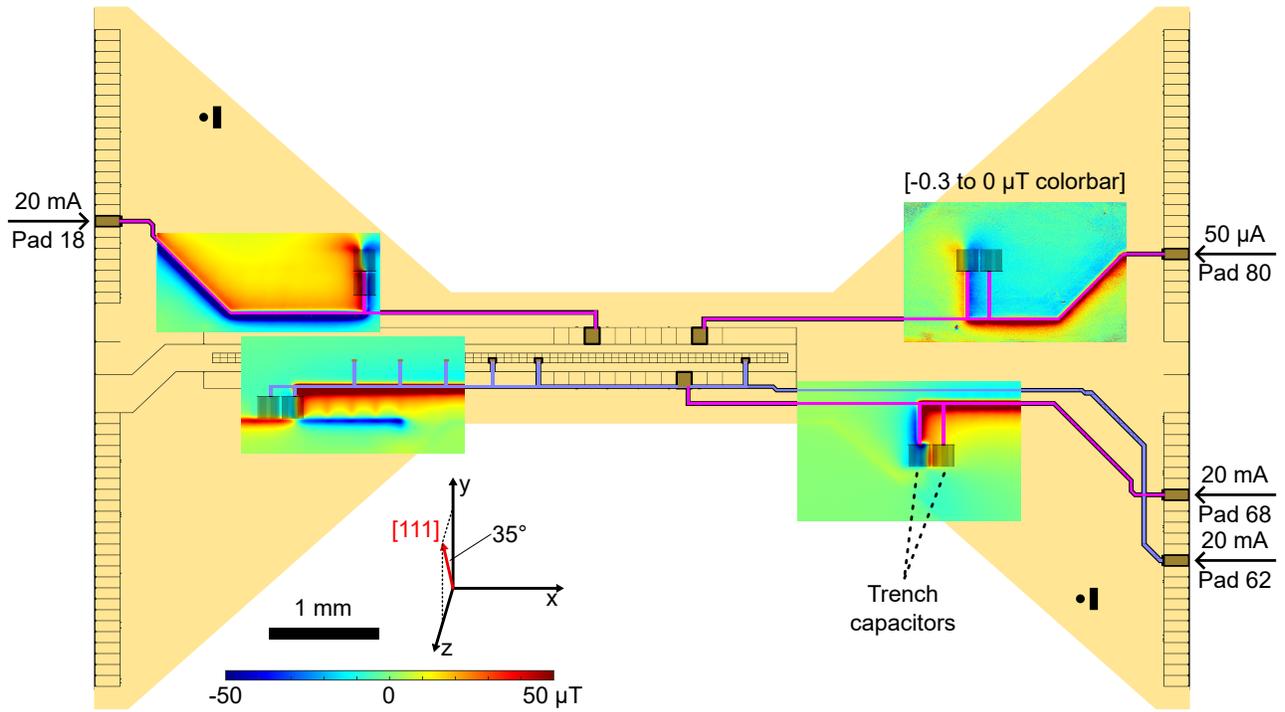

**Supplementary Figure** S1.  Magnetic image of current in three low-resistance shorts (Pads 18, 62, and 68) and high-resistance short (Pad 80) of a Perigrine trap.  The overlays show shorts to ground at trench capacitors, which is consistent with what was observed in the Roadrunner ion trap chip in the main text.

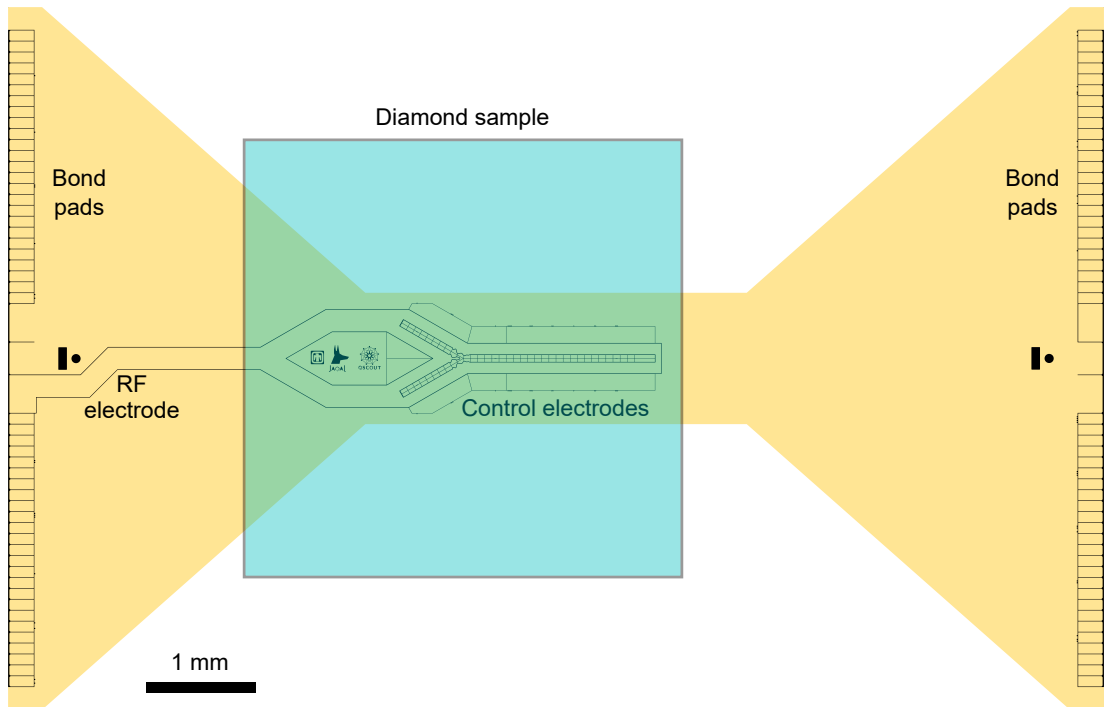

**Supplementary Figure** S2.  The Roadrunner bowtie top metal layer, showing the bond pads, control electrodes, and RF electrode.  The diamond sample is placed on top for MFI experiments, shown here at an example position on the bowtie.



\end{document}